\documentclass[10pt, a4paper]{article}
\usepackage{authblk}
\usepackage[normalem]{ulem}
\usepackage{dsfont}
\usepackage{amsmath, amssymb}
\usepackage{xfp}
\usepackage[margin=1in]{geometry}
\usepackage{cite}

\usepackage{diagbox}
\usepackage{graphicx}
\graphicspath{{final_figs/}}
\usepackage{cite}
\usepackage[table]{xcolor}

\usepackage{color}
\usepackage{hyperref}
\hypersetup{
colorlinks=false}
\usepackage{multirow}

\usepackage{todonotes}
\presetkeys{todonotes}{backgroundcolor=yellow}{}

\def\<{\langle}
\def\>{\rangle}

\newcommand{\eval}[1]{\mathbf E\left( #1 \right)}

\newcommand{\mytodo}[1]{}

\newcommand{\prob}[1]{{\ensuremath{\mathbf P\left( #1 \right)}}}

\title{Estimation of heavy tails in optical non-linear processes}

\author[1,2]{\'Eva R\'acz}
\author[1]{L\'aszl\'o Ruppert}
\author[1]{Radim Filip}

\affil[1]{{\small Department of Optics, Palacky University, 17. listopadu 12, 771 46 Olomouc, Czech Republic}}
\affil[2]{{\small Department of Theoretical Physics, Budapest University of Technology and Economics, M\H{u}egyetem rkp. 1., 1111 Budapest, Hungary}}

\date{}
\begin{document}
\maketitle


\begin{abstract}
In optical non-linear processes rogue waves can be observed, which can be mathematically described by heavy-tailed distributions. These distributions are special due to the fact that the probability of registering extremely high intensities is significantly higher than for the exponential distribution, which is most commonly observed in statistical and quantum optics. 
 The current manuscript gives a practical overview of the generic statistics toolkit concerning heavy-tailed distributions and proposes methods to deal with issues specific to non-linear optics. We take a closer look at supercontinuum  generation, where rogue waves were already observed. We propose modifications to the Hill estimator to deal with detector saturation as well as corrections introduced by pumping the process by bright squeezed vacuum. The suggested methodology facilitates statistically reliable observation of heavy-tailed distribution in non-linear optics, nanooptics, atomic, solid-state processes and optomechanics.    
\end{abstract}



\section{Introduction} 

Heavy-tailed distributions, and power-law (or Pareto) distributions in particular have been reported from a very broad range of areas, including earthquake intensities \cite{Gutenberg1944,Christensen2002,Newberry2019}, avalanche sizes \cite{Birkeland2002}, solar flares \cite{Lu1991}, degree distributions of various social and biological networks \cite{Pastor2001,Stumpf2005,Newman2005}, incomes \cite{Pareto1896,Yakovenko2009}, insurance claims \cite{Shpilberg1977,Rootzen1995}, number of citations of scientific publications \cite{Price1965,Redner1998,Golosovsky2017}, and many more. For financial institutions, the importance of heavy-tailed behavior comes from the fact that a simple Gaussian model severely underestimates the risks associated with different products or investment strategies, which in turn results in considerable losses. This is why the mathematical background of heavy-tailed distributions and their estimation have been most extensively studied in this context (\hspace{1sp}\cite{Mandelbrot1960, Mandelbrot1963,Rachev2003,Embrechts1999} and others). 

For physicists, in turn, the general fascination with power laws comes from the concept of universality classes in statistical physics, which describe the behavior of various systems close to their critical points, irrespective of the details of the mechanisms of the systems \cite{Wilson1983}. Inspired by this, and using the increasing availability of good-quality data sets, many of the field have ventured into interdisciplinary waters where power-law behavior has been reported; thereby creating entirely new disciplines like econophysics \cite{Kertesz1997,Schinckus2016}. In such investigations, the emphasis has been mostly on understanding the emergence of power-law behavior \cite{Bak1991, Albert2002, Farmer2004b}. In contrast, the problem of practical estimation methods of the exponent associated with the power-law (or indeed the verification of power-law behavior) has received relatively little attention from the community. Notable exceptions are \cite{Newman2005, Clauset2009}, in which the power-law nature of numerous phenomena has been questioned. 

Turning to non-linear optics \cite{Boyd2008, Grynberg2010}, the problem formulation is, however, somewhat different. Optical experiments producing light with heavy-tail intensity distributions allow high repetition rates and large samples to study unstable non-linear phenomena and their sources \cite{Barthelemy2008, Mercadier2009}. Moreover, unlike in social or financial contexts, experiments can be repeated. For intensive pumps, non-linear systems can produce bright signals with directly measurable intensity fluctuations. In statistical optics \cite{Goodman1985}, coherence theory and quantum optics \cite{Mandel1995}, the probability density function of intensities is a subject of investigation for intensive light beams. Remarkably, it allows a direct observation of macroscopic quantum phenomena \cite{Iskhakov2009, Iskhakov2011, Iskhakov2012, Iskhakov2016a, Iskhakov2016b}. Intensity distributions in supercontinuum generation setups \cite{Solli2007,Ruban2010,Akhmediev2016} being heavy-tailed -- hence the term \emph{rogue waves} -- does have a theoretical basis. 
So our aim is not to decide whether the observed distribution is heavy-tailed or not. We rather propose ways to estimate the \emph{tail exponent} of the distribution in the presence of experimental imperfections,  which can help in experimental control and design. The issue of detector saturation, for example, should be paid special attention to since for intensity distributions with extremely heavy tails, it cannot be solved by a simple re-calibration. There is always going to be some portion of the data that will be affected by saturation. Low-intensity statistics, on the other hand, are mostly determined by background noise. Consequently, there is only a limited interval of intensities within which the observations can be used for estimation purposes and this hugely affects the efficiency of any estimation procedure. If we are unlucky, the intervals affected by the background noise and detector saturation overlap, and the experiment cannot be salvaged. If, however, there is a portion that is useful, we propose ways to gain an initial estimate of the tail exponent (and by extension, higher quantiles) which helps design further measurement setups. The procedures presented in the current manuscript can be used to quantify the instabilities resulting in extreme events common in optical non-linear processes \cite{Manceau2019,Spasibko2017}, and potentially also in non-linear optomechanics \cite{Brawley2016,Siler2018}, four-wave mixing in atomic vapors \cite{McCormick2007, Guerrero2020} and wave-mixing processes in superconducting circuits \cite{Aban2004, Sivak2019, Mundhada2019}. Beyond fundamental interest, extreme events can be used to produce highly non-classical effects distillable to large squeezing and entanglement for quantum technology \cite{Heersink2006,Dong2008}.  In parallel, they also inspire an investigation of optical Maxwell demon principle \cite{Vidrighin2016} for heavy-tailed distributions.  

The article is structured as follows: In Section \ref{sec:background}, we give a brief overview of the mathematical background and clarify terminology. Section \ref{sec:tools} describes a generic estimation toolkit with pointers to more sophisticated methods. Section \ref{sec:intensity} proposes variations of the generic methods for evaluating numerical data in non-linear optics specifically.


\section{Background and terminology}\label{sec:background}

Let us first provide an overview of the terminology and concepts concerning \emph{heavy-tailed distributions}, used throughout this work.

In general, distributions that decay slower than exponential are referred to as heavy-tailed. To give a clear mathematical formulation of this concept, it is useful to introduce the \emph{tail function} (also referred to as survival function, or complementary distribution function), defined for an arbitrary real-valued random variable \(X\) as
\[\overline F(x) \equiv  \prob{X \geq x}.\] 
In other words, this function describes the probability that the variable reaches or exceeds a threshold \(x\). It is related to the more familiar distribution function (DF) \(F(x)\) through \(\overline F (x) = 1 - F(x)\), and to the probability density function (PDF) \(f(x)\) through \(\overline F(x) = \int_x^{\infty} f(u)\, \mathrm d u\). 
In what follows, we will only consider the right tail of distributions, that is, the behavior of the largest values, and for the sake of simplicity, suppose that we are dealing with positive-valued random variables like optical intensities. The whole treatment can be straightforwardly extended to the left tails of distributions.

Using the tail function (TF), a \emph{heavy-tailed distribution} can be defined as a distribution for which
\begin{equation}\label{heavy_def}
C \equiv \mathop{\underline{\lim}}_{x\to\infty}\frac{-\ln \overline F(x)}{x} = 0.
\end{equation}
There are other equivalent definitions \cite{HeavyTextBook}, we prefer the latter formulation since it is the one which is most in line with the somewhat vague notion of an ``L-shaped'' distribution used in connection with rogue waves (\hspace{1sp}\cite{ShapingLight}, for example). Note that if the limit $C$ in \eqref{heavy_def} is a finite positive value, the distribution decays asymptotically at an exponential rate, while if it is infinity, the distribution decays faster than exponential. Definition \eqref{heavy_def} includes even distributions whose moments of any order are finite, for example the log-normal distribution and the Weibull distribution with a shape parameter lower than one \cite{Bohm2010_IntroToStatistics}.

\emph{Pareto-type distributions} (or regularly-varying distributions) form a subset of heavy-tailed distributions and are defined as
\begin{equation}
\label{eq:RV}
\overline F(x) = x^{-\alpha}\cdot L(x),
\end{equation}
with \(L(x)\) being a slowly-varying function (for any \(t > 0\), \(\lim_{x\to\infty}L(tx)/L(x) = 1\); slowly-varying functions include for example the logarithm function and functions that have a finite limit in infinity) and $\alpha > 0$. For the Pareto distribution, $L(x) = \mathrm{const}$, corresponding exact power-law behavior. The exponent \(\alpha\) is referred to as the \emph{tail exponent}. Note that this exponent describes the decay of the tail function, the exponent corresponding to the decay of the PDF is \(\alpha+1\). The moments \(\eval{X^a}\) are finite for all $a < \alpha$ and infinite for all $a > \alpha$. Whether $\eval{X^\alpha}$ itself is finite depends on the function \(L(x)\).

What makes heavy-tailed distributions statistically special is the fact that many traditional procedures based on the mean and the variance are inapplicable if the first and second moments do not exist. It is possible, of course, to compute sample averages, but they are meaningless if the corresponding expected value is not finite for the underlying distribution. That is, as the number of observations is increased, such sample averages do not converge to a finite number. The lack of definite moments limits many evaluations in statistical optics \cite{Goodman1985}, coherence optics and quantum optics \cite{Mandel1995}. One should not, for example, calculate second-order quantities like correlation \cite{Boitier2011, Spasibko2017, Zhou2017, Zhang2019} for $\alpha < 2$. Furthermore, the traditional central limit theorem, which lies at the heart of many applied models, is not applicable, either.

The question what can be said about the largest observations for such distributions arises quite naturally. There are limit laws concerning the maxima of samples and the distribution of threshold exceedances. Very simply stated, if a distribution has the form \eqref{eq:RV}, the distribution of sample maxima tends to an extreme value distribution described by the DF $\exp\left\{-(1+\gamma x)^{-1/\gamma}\right\}$, with $0 < \gamma = \alpha^{-1} < \infty$. The distribution of the exceedances of a sufficiently large threshold $l$ (that is, the random variable $X-l \mid X > l$) tends in turn to a generalized Pareto distribution with DF $1 - (1+\gamma x)^{-1/\gamma}$, $\gamma = \alpha^{-1}$. Both limit laws hold, of course, given proper normalizing constants, for details see \cite{Beirlant_StatisticsOfExtremes}. These limit laws can be straightforwardly used to model the behavior of the underlying distribution beyond the largest observed value. 

\begin{figure}[t]
\centering
\newcommand\shift{-0.03096769}
{
\begingroup%
  \makeatletter%
  \providecommand\color[2][]{%
    \errmessage{(Inkscape) Color is used for the text in Inkscape, but the package 'color.sty' is not loaded}%
    \renewcommand\color[2][]{}%
  }%
  \providecommand\transparent[1]{%
    \errmessage{(Inkscape) Transparency is used (non-zero) for the text in Inkscape, but the package 'transparent.sty' is not loaded}%
    \renewcommand\transparent[1]{}%
  }%
  \providecommand\rotatebox[2]{#2}%
  \newcommand*\fsize{\dimexpr\f@size pt\relax}%
  \newcommand*\lineheight[1]{\fontsize{\fsize}{#1\fsize}\selectfont}%
  \ifx\svgwidth\undefined%
    \setlength{\unitlength}{351.50464876bp}%
    \ifx\svgscale\undefined%
      \relax%
    \else%
      \setlength{\unitlength}{\unitlength * \real{\svgscale}}%
    \fi%
  \else%
    \setlength{\unitlength}{\svgwidth}%
  \fi%
  \global\let\svgwidth\undefined%
  \global\let\svgscale\undefined%
  \makeatother%
  \begin{picture}(1,0.67871066)%
    \lineheight{1}%
    \setlength\tabcolsep{0pt}%
    \put(0,0){\includegraphics[width=\unitlength]{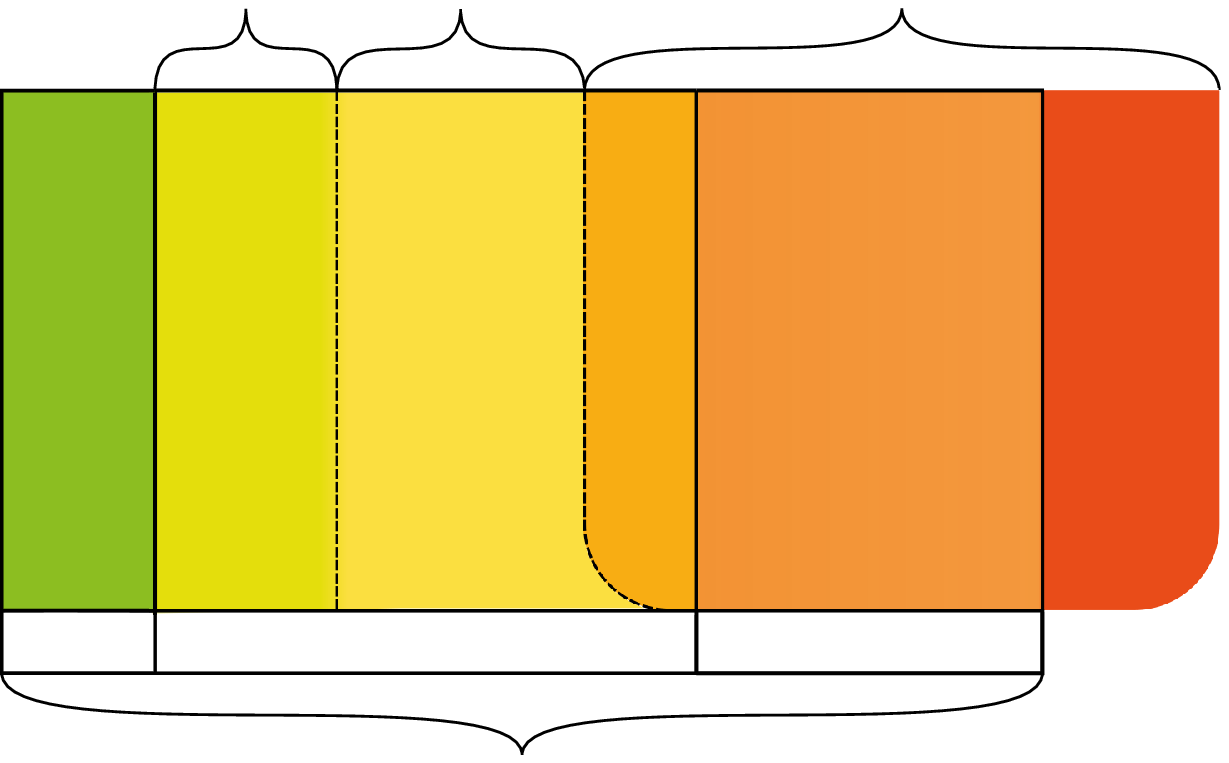}}%
    \put(0.34749046,\fpeval{0.36512741+\shift}){\color[rgb]{0,0,0}\rotatebox{90}{\makebox(0,0)[t]{\lineheight{1.25}\smash{\begin{tabular}[t]{c}Exponential \cite{Manceau2019} \\\\Gamma \cite{Manceau2019}\end{tabular}}}}}%
    \put(0.9348413,\fpeval{0.25561001 + \shift}){\color[rgb]{0,0,0}\rotatebox{90}{\makebox(0,0)[lt]{\lineheight{1.25}\smash{\begin{tabular}[t]{l}Log-Pareto \cite{Cormann2009}\end{tabular}}}}}%
    \put(0.63832751,\fpeval{0.11596769+\shift}){\color[rgb]{0,0,0}\makebox(0,0)[lt]{\lineheight{1.25}\smash{\begin{tabular}[t]{l}$\alpha = \gamma^{-1} > 0$\end{tabular}}}}%
    \put(0.3072914,\fpeval{0.11596769 + \shift}){\color[rgb]{0,0,0}\makebox(0,0)[lt]{\lineheight{1.25}\smash{\begin{tabular}[t]{l}$\gamma = 0$\end{tabular}}}}%
    \put(0.02203698, \fpeval{0.11596769 + \shift}){\color[rgb]{0,0,0}\makebox(0,0)[lt]{\lineheight{1.25}\smash{\begin{tabular}[t]{l}$\gamma < 0$\end{tabular}}}}%
    \put(0.23445492, \fpeval{0.00048338 + \shift}){\color[rgb]{0,0,0}\makebox(0,0)[lt]{\lineheight{1.25}\smash{\begin{tabular}[t]{l}Max domain of attraction\end{tabular}}}}%
    \put(0.53854442, \fpeval{0.25299836 + \shift}){\color[rgb]{0,0,0}\rotatebox{90}{\makebox(0,0)[lt]{\lineheight{1.25}\smash{\begin{tabular}[t]{l}Log-normal \cite{Milonni2004}\end{tabular}}}}}%
    \put(0.88516223, \fpeval{0.11596769 + \shift}){\color[rgb]{0,0,0}\makebox(0,0)[lt]{\lineheight{1.25}\smash{\begin{tabular}[t]{l}$\alpha = 0$\end{tabular}}}}%
    \put(0.73903773, \fpeval{0.65277305 + \shift}){\color[rgb]{0,0,0}\makebox(0,0)[t]{\lineheight{1.25}\smash{\begin{tabular}[t]{c}Heavy-tailed: $C= 0$\end{tabular}}}}%
    \put(0.07644239, \fpeval{0.20518685 + \shift}){\color[rgb]{0,0,0}\rotatebox{90}{\makebox(0,0)[lt]{\lineheight{1.25}\smash{\begin{tabular}[t]{l}Uniform \cite{Jaynes1957}, Beta \cite{Dmitruk2009}\end{tabular}}}}}%
    \put(0.28636368, \fpeval{0.65277305 + \shift}){\color[rgb]{0,0,0}\makebox(0,0)[lt]{\lineheight{1.25}\smash{\begin{tabular}[t]{l}$0 < C < \infty$\end{tabular}}}}%
    \put(0.14326637, \fpeval{0.65277305 + \shift}){\color[rgb]{0,0,0}\makebox(0,0)[lt]{\lineheight{1.25}\smash{\begin{tabular}[t]{l}$C = \infty$\end{tabular}}}}%
    \put(0.21421823, \fpeval{0.26865091 + \shift}){\color[rgb]{0,0,0}\rotatebox{90}{\makebox(0,0)[lt]{\lineheight{1.25}\smash{\begin{tabular}[t]{l}Gaussian \cite{Jaynes1957} \end{tabular}}}}}%
    \put(0.63825478, \fpeval{0.47442825 + \shift}){\color[rgb]{0,0,0}\makebox(0,0)[lt]{\lineheight{1.25}\smash{\begin{tabular}[t]{l}Pareto \cite{Pareto1896}\end{tabular}}}}%
    \put(0.63106856, \fpeval{0.39677595 + \shift}){\color[rgb]{0,0,0}\makebox(0,0)[lt]{\lineheight{1.25}\smash{\begin{tabular}[t]{l}Cauchy \cite{Svelto2010}\end{tabular}}}}%
    \put(0.65162017,\fpeval{0.31306397 + \shift}){\color[rgb]{0,0,0}\makebox(0,0)[lt]{\lineheight{1.25}\smash{\begin{tabular}[t]{l}Levy \cite{Lepri2007}\end{tabular}}}}%
    \put(0.60042899, \fpeval{0.23599199 + \shift}){\color[rgb]{0,0,0}\makebox(0,0)[lt]{\lineheight{1.25}\smash{\begin{tabular}[t]{l}Log-gamma \cite{Ghitani2018}\end{tabular}}}}%
  \end{picture}%
\endgroup%
}
\vspace*{5mm}
\caption{Categorization of univariate continuous distributions according to the properties of their right tails: speed of decay compared to exponential ($C$), tail exponent ($\alpha$), and behavior of maxima ($\gamma$); including a few examples for each category.{\nocite{Manceau2019,Jaynes1957,Dmitruk2009,Milonni2004,Svelto2010,Lepri2007,Cormann2009, Ghitani2018}} \mytodo{Add more references, include log-gamma.}}\label{fig:venn}
\end{figure}

There are heavy-tailed distributions whose tails are heavier than \eqref{eq:RV} (for example the log-Pareto constructed by exponentiating a Pareto-distributed variable), these limit laws do not apply to them. There are also heavy-tailed distributions whose tail is lighter than \eqref{eq:RV}, for these the DF of the maxima tends to $e^{-e^{-x}}$, whereas threshold exceedances tend to an exponential distribution; these distributions arise as the natural limit for $\gamma = 0$. Distributions with a light, but infinite tail also belong to the $\gamma=0$ domain of attraction. The limit distribution of the maxima of random variables with a finite upper bound is also $\exp\left\{-(1+\gamma x)^{-1/\gamma}\right\}$, but with $\gamma < 0$. Figure \ref{fig:venn} summarizes this categorization according to tail heaviness.

In what follows, we will deal with the extreme value index \(\gamma\) instead of the tail exponent \(\alpha = \gamma^{-1}\). This has two practical reasons: firstly, \(\gamma\) is the quantity that is used in all mathematical publications, so the quite extensive mathematical theory is based on that quantity; and secondly, in the context of supercontinuum generation, \(\gamma\) is the quantity proportional to the mean intensity of the pump, so it is in a way more handy than \(\alpha\).


\section{Basic tools for investigating power-law tails}\label{sec:tools}
The purpose of this section is to show physicists some simple, visual tools for assessing power-law behavior and also give references on improving the behavior of the estimators. The reason why we have not picked a single favorite estimator is that in order to have a reliable assessment of power-law behavior, it is better to use more than one tool and see whether they produce consistent results.

\emph{Examples:} For demonstration purposes we will use in the next sections computer-generated samples of different distributions; their properties are summarized in Table \ref{tab:dist1}. The size of the generated samples was $10^4$. We did not use solely regularly-varying distributions \eqref{eq:RV} because we would also like to show what the tools produce when used with distributions that do not have a power-law tail. The exponential distribution is a standard thin-tailed distribution, compared to which the heavy-tailed property itself is defined. The log-normal distribution is, according to the definition \eqref{heavy_def} heavy-tailed, but it does not have a finite tail exponent. A log-gamma distributed variable can be created by exponentiating a gamma-distributed variable the same way one transforms a normally distributed variable into a log-normal; it is an example for a regularly-varying distribution with \(L(x) \neq \mathrm{const}\). 
The Pareto distribution corresponds to pure power-law behavior, and is used for demonstrating the best-case scenario (it is also the \(a = 1\) special case of the log-gamma distribution).

\begin{table}[tbh]
\centering
\renewcommand{\arraystretch}{2}
\begin{tabular}{c|c|c|c|c}
Name & PDF & Heavy-tailed? & \(\gamma\) \\
\hline
Exponential & \(\displaystyle \lambda e^{-\lambda x}\) & N & 0 \\
Log-normal & 
\(\displaystyle \frac 1{\sqrt{2\pi \sigma^2}} \frac 1 x \exp\left\{-\frac{\ln^2 x^2}{2\sigma^2}\right\}\) 
& Y & 0 \\
Log-gamma &
\(\displaystyle 
\frac{b^a}{\Gamma(a)} (\ln x)^{a-1} \cdot x^{-b -1} 
\) &  Y & \(b^{-1}\)\\
Pareto  & \(\displaystyle \alpha\cdot x ^{-\alpha-1}\) & Y & \(\alpha^{-1}\) 
\end{tabular}
\caption{Distributions used for demonstration purposes.} \label{tab:dist1}
\end{table}

\subsection{Histogram}

Preparing the histogram is probably the most wide-spread way to visualize random samples, and is definitely go-to tool for many physicists. This is why we devote more attention to it in this paper than it deserves from the mathematical point of view. It involves defining a discrete set of bins over the number line and then counting how many observations of the random variable fall in each bin. Given a linear set of bins, the histogram provides an estimate of the PDF of the underlying distribution. If this distribution is, at least approximately, Pareto, then the histogram should be linear on a log-log plot with the absolute value of the slope equal to the tail exponent plus one. One can even perform a least-squares fit to obtain the slope of the line to estimate the exponent. There are, however, three major problems with this approach:
\begin{enumerate}
\item Due to the power-law nature, there will be only a few observations (if any) on the right end of a linear set of bins, meaning that exactly for large values, where the power-law property itself should be more pronounced, there will be considerable variance.
\item As taking the logarithm and the expectation are not interchangeable, the expected value of the logarithm of the frequencies is not equal to the logarithm of the PDF. Furthermore, the variance of the frequencies depends on the location. So the basic prerequisites of an ordinary least squares fit do not hold.
\item The choice of bins has a marked effect on the outcome.
\end{enumerate}

\begin{figure}[ptb!]
\centering
\includegraphics[width=\textwidth]{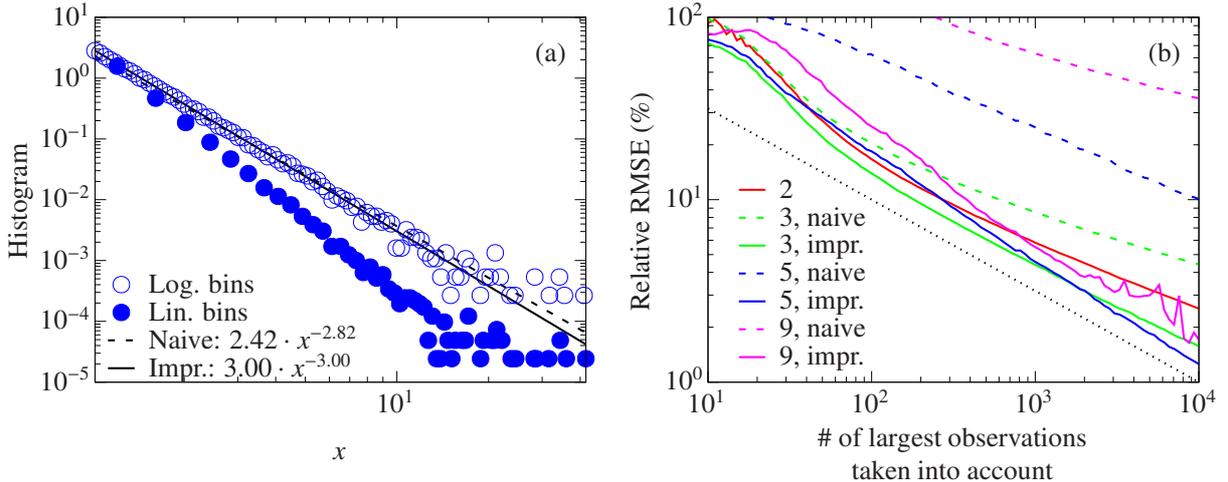}
\caption{(a) Linearly (filled circles) and logarithmically binned (empty circles) histograms for a single sample of Pareto (\(\alpha = 3\)) data, sample size $10^5$, 100 bins. The dashed line shows the result of the naive estimator, the solid line shows the result of the improved version. For the linearly binned histogram, the analytic formula is \(3\cdot x^{-4}\), for the logarithmically binned \(3\cdot x^{-3}\). (b) Relative root mean squared error (RMSE) of the naive histogram estimator of \(\alpha^{-1}\) and the improved version as a function of how many of the largest observations were taken into account. Values were calculated using $10^4$ Pareto-distributed samples of size \(10^4\). The dotted black line corresponds to the Cramér--Rao bound for the Pareto distribution, \(1/\sqrt{k}\); the colored dashed lines show the performance of the naive estimator; and the colored solid lines correspond to the improved estimator. The different colors indicate how many bins were used to construct the histogram (see the legend).}\label{fig:hist_est}
\end{figure}

To help with the first problem, one can use logarithmic bins instead of linear ones. However, it is important to be aware of the fact that using logarithmic bins, that is, bin sizes that are linear on a logarithmic scale is equivalent to preparing a linearly binned histogram of $Y=\ln X$. Consequently, this version of the histogram approximates the probability density function of $Y$, $g(y)$. This change of variables results in \(g(\ln x)=x \cdot f(x)\), meaning that the slope of the linearly binned histogram is $\alpha+1$, while the slope of the logarithmically binned version is $\alpha$ instead, see figure \ref{fig:hist_est}(a). 

Concerning the second point, if one does not take the logarithm of the frequencies, there is no problem with expectations, and the variances can be calculated explicitly. A consistent version of the histogram approach is preparing the histogram of $Y = \ln X$, and performing a weighted least-squares fit of $A\cdot e^{-By}$. The weights are needed to make the data homoscedastic \cite{Aitken1934}, and can be calculated as $\left[\hat p_k(1 - \hat p_k)\right]^{-1}$, with $\hat p_k$ denoting the fraction of observations within the $k^\mathrm{th}$ bin. The choice of bins in this setting is especially problematic, since empty bins should to be avoided (or, alternatively, discarded when performing the fit). 

The third issue cannot be completely eliminated, however, figure \ref{fig:hist_est}(b) shows the results of a small-scale simulation experiment based on purely power-law samples (corresponding to the best-case scenario). We have taken \(N_{\mathrm{samples}} = 10^4\) samples, each consisting of $N=10^4$ elements. For each sample, for \(k = 1, 2, \ldots,N\) we have calculated the values of the estimators based on the top \(k\) observations and compared them to the known value of the extreme value index \(\gamma\) to obtain
\[\mathrm{Rel.\ RMSE}_m(k) \equiv \frac{1}{\gamma}\cdot 
\sqrt{\frac 1 {N_{\mathrm{samples}}} 
\sum_{i = 1}^{N_{\mathrm{samples}}} \left(\hat\gamma_{m,i}(k) - \gamma\right)^2,}
\]
with \(m\) denoting the method used. This means that for a given value of \(k\), we prepared 4 histograms (2, 3, 5, and 9 bins), and calculated the estimate using the naive (least-squares linear fit on log-log scale) and the improved version (weighted least squares exponential fit on lin-log scale) of the estimator as well for either. First of all, the simulation showed how bad the performance of the naive histogram estimator really is. Secondly, and probably a little counter-intuitively, it showed how few bins are required in order to minimize the error of either version of the estimator based on the histogram compared to how many bins one would use for visualization. 
Considering, however, that one only needs to estimate a single parameter, increasing the number of points does not necessarily help if these in turn become less accurate.
To get an estimate of how many bins should be used for a given number of observations \(k\), one can consider the Rényi representation theorem \cite{Renyi1953}. Using it, it is straightforward to show that the difference between the log-spacing between the largest and smallest element of a purely Pareto-distributed sample of size \(k\) is \(\alpha^{-1}\cdot[1 + 1/2 + \ldots + 1/(k-1)] \approx \alpha^{-1}[\gamma^* + \ln(k-1)]\), with \(\gamma^* \approx 0.5772\) the Euler--Mascheroni constant. Furthermore, supposing that there is an ideal (log) bin width that minimizes the RMSE of a histogram estimator independently of the sample size, it has to be proportional to \(\alpha^{-1}\) (since that is the scale parameter of the logarithmically transformed sample), so the total number of bins should be a linear function of \(\ln(k-1)\). Based on our simulations, for \(k=100\), 3 bins are ideal, for \(k=10^4\) 5 bins produce the best results using the improved version of the histogram estimator (see figure \ref{fig:hist_est}(b)). Interestingly, for the naive version it seems to be always better to prepare just 2 bins.

All in all, the histogram need not be completely discarded as a tool even when working with heavy-tailed distributions, provided that certain changes are applied to the naive estimator: use logarithmic binning, use only a few bins (if the histogram resembles a broomstick-like in figure \ref{fig:hist_est}(a), reduce the number of bins), and use weighted least-squares to fit an exponential to the logarithmically transformed data. Nevertheless, the histogram remains first and foremost a visual tool in the heavy-tailed context.


\subsection{Empirical tail function and QQ estimator}\label{sec:TF}

\begin{figure}[tb]
\centering
\includegraphics[width=0.9\textwidth]{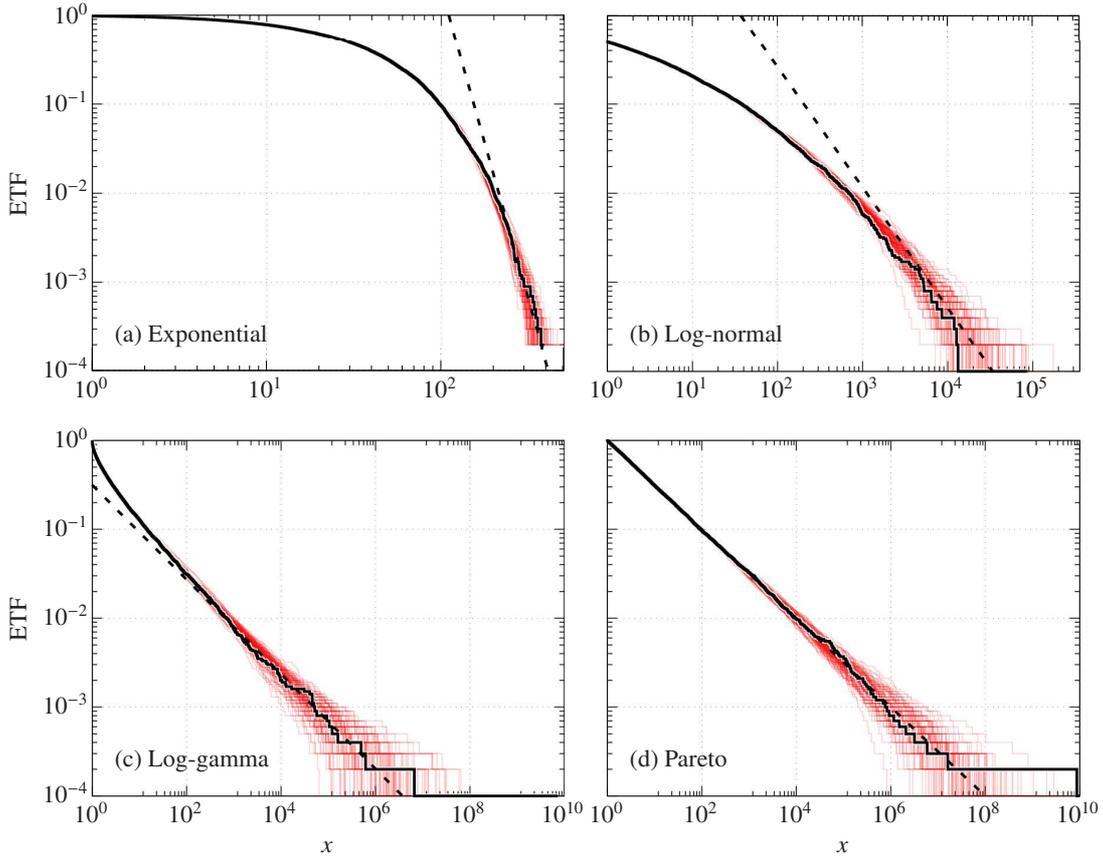}
	\caption{Empirical tail functions for 100 samples of length \(10^4\), for the distributions: (a) Exponential (\(\lambda = 0.023\)), (b) Log-normal (\(\sigma = 2.8\)), (c) Log-gamma (\(a = b = 0.5\)), and (d) Pareto \((\alpha = 0.5)\). One of the ETF-s for each distribution is shown in black, the others in red. The dashed lines correspond to the tangent line to the analytic TF for the largest values. The slopes of the dashed lines are (a) -1.35, (b) -0.53, and (c) -0.5. Note that distinguishing a log-normal sample from a regularly varying one is non-trivial since the tangent of the ETF based on a finite sample is never vertical.}\label{fig:etfhist}
\end{figure}

The empirical tail function (ETF) presents a simple alternative to histograms as a visual tool that does not depend on an arbitrary binning procedure. Given a sample of independent, identically distributed (iid) observations, \(\left\{x_1, x_2,\ldots, x_N\right\}\), sorted in descending order \(x_{(1)}\geq x_{(2)}\geq\ldots \geq x_{(N)}\), the empirical tail function is given as
\begin{equation}
\overline F^*\left(x_{(k)}\right) = \frac k N,
\label{eq:emp.cdf}
\end{equation}
or equivalently, for an arbitrary threshold \(l\in \mathbb R\),
\[\overline F^*(l) = \frac 1 N \sum_{i = 1}^N \mathds 1\left\{x_i \geq l\right\},\]
with \(\mathds 1 \{\cdot\}\) denoting the indicator function. In other words, one has to check what proportion of the observations exceed the limit \(l\); \(l = x_{(k)}\) yields the first definition, equation \eqref{eq:emp.cdf}. If the sample is power-law distributed (at least for the largest observations) the ETF should be linear on a log-log plot (again, for the largest observations), with the slope equal to \(-\alpha\) (see figure \ref{fig:etfhist}). Since the TF is the integral of the PDF, the ETF is considerably smoother than the histogram, which makes it easier to detect deviations from power-law behavior visually. 

As for a numerical estimate of \(\alpha\), it is, of course, possible to perform a least-squares fit on \(\left\{\ln x_{(k)}, \ln\frac k N \right\}\) (and that also yields much better results than the naive histogram approach), however, it is better to do this with the roles reversed, that is, by treating \(\ln \frac k N\) as the independent variable. The latter procedure is referred to as the QQ estimator \cite{Kratz1996}, with QQ standing for quantile-quantile. The essential difference is that in the ETF version, one divides by the sample average \(\left\langle\ln^2 x_{(k)}\right\rangle - \left\langle\ln x_{(k)}\right\rangle^2\), and in the QQ version a deterministic number, \(\left\langle\ln^2 \frac{k}{N}\right\rangle - \left\langle\ln \frac{k}N\right\rangle^2\). The QQ version is therefore more stable.

In general, a QQ plot is constructed by plotting the sorted observations against the matching quantiles of the theoretical distribution. If the underlying distribution matches the theoretical distribution (up to a linear transformation), then the QQ plot should be linear. In the specific case of a Pareto-distributed variable, one transforms it into an exponentially distributed variable by taking the logarithm and plots the sample as a function of the standard exponential quantiles, that is, the plot consists of the points $\left\{ -\ln \frac k N, \ln x_{(k)}\right\}$ and the slope of the line is \(\alpha^{-1}\). In essence, this corresponds to exchanging the two axes of the ETF plot on a log-log scale. The QQ estimator of \(\gamma = \alpha^{-1}\) is then calculated as the slope of the LS fit on the QQ plot. 

This is a very simple estimator, and it has been shown to be weakly consistent \cite{Resnick_Heavy-TailPhenomena}. However, the issues present in the naive histogram estimator are present here, too. Namely the expected value of $\ln X_{(k)}$ is not $-\alpha^{-1}\cdot \ln \frac k N$, and the variance of $\ln X_{(k)}$ depends on $k$. What is more, $X_{(i)}$ and $X_{(j)}$ are not independent. The problem has been addressed in \cite{Aban2004} for purely Pareto distributed samples, taking into account both the issue of expectations and the covariance matrix of order statistics. The authors show that the solution of the generalized regression problem is equivalent to the Hill estimator discussed in section \ref{sec:hill}, which should not come as a surprise for since it is the maximum likelihood estimator. The same issue has been addressed in \cite{Beirlant1999}, under more general assumptions on the underlying distribution, yielding an improved regression estimator. Nevertheless, even the basic QQ estimator is also a viable, although usually inferior, alternative to the Hill estimator.

\subsection{Hill plot}\label{sec:hill}

The Hill estimator can be obtained as the conditional maximum likelihood estimator of the reciprocal of the tail exponent \(\alpha^{-1}\) for Pareto-distributed data \cite{hill1975}:  
\begin{equation}
\widehat{\alpha^{-1}}(k) \equiv \hat\gamma^{\mathrm H}(k) = \frac 1 {k}\sum_{i = 1}^{k} \ln x_{(i)} - \ln x_{(k+1)},
\label{eq:hill}
\end{equation}
with \(x_{(i)}\), \(i = 1, 2, \ldots, N\) denoting the \(i^{\mathrm{th}}\) largest element of the sample. In a more general setting, given a sufficiently large number of observations from a regularly-varying distribution, it converges to the extreme value index of the distribution (the rate of convergence depends on the distribution) \cite{beirlant2005}. For pure Pareto\((\alpha)\) data, \(\eval{\hat \gamma^{\mathrm H}(k)} = \alpha^{-1}\), and \(\mathrm{RMSE}\left({\hat \gamma^{\mathrm H}(k)}\right) = \alpha^{-1}/\sqrt{k}\).  

\begin{figure}[tb]
\centering
\includegraphics[width=0.9\textwidth]{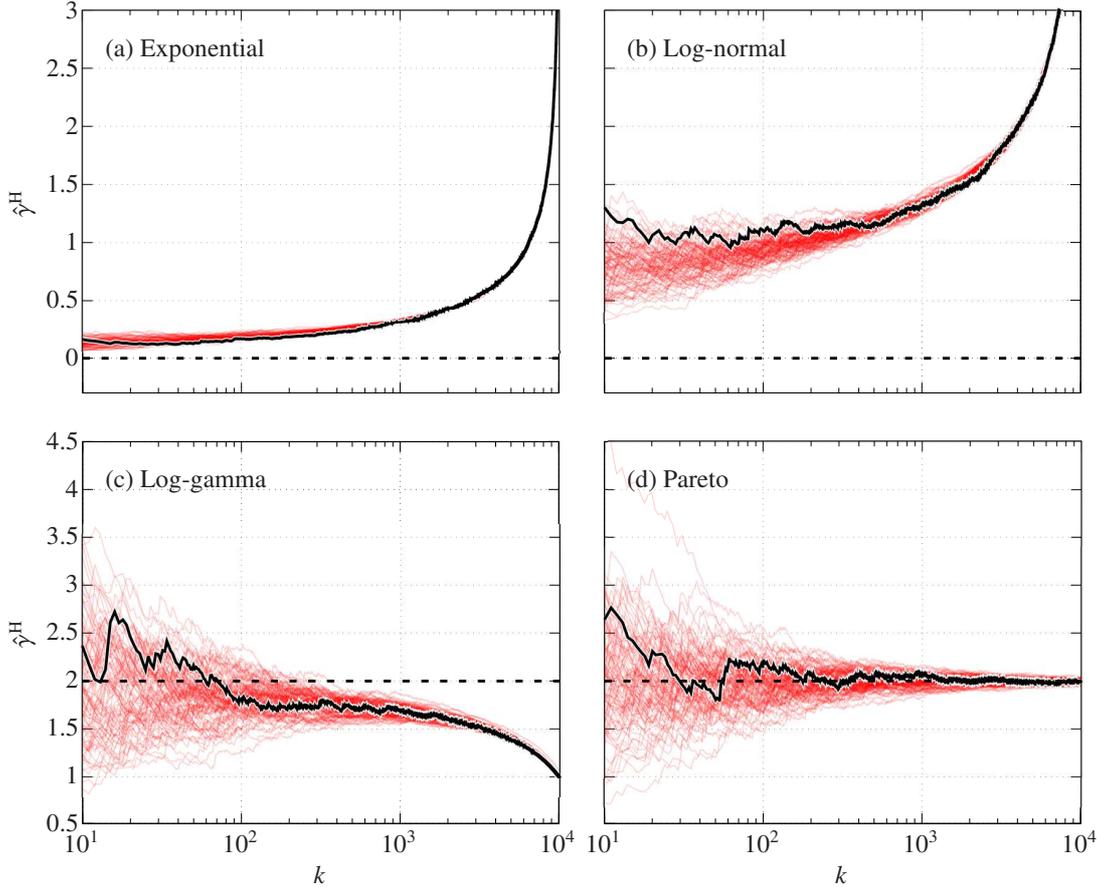}
\caption{Hill plot for 100 computer-generated data sets of length $10^4$ for different distributions. The plot for a single realization is highlighted in black for each distribution. The parameters of the distributions were the following: (a) Exponential (\(\lambda = 0.023\)), (b) Log-normal \(\mu = 0, \sigma = 2.8\), (c) Log-gamma: \(a = b = 0.5\), and (d) Pareto: \(\alpha = 0.5\). The true values for the extreme value index \(\gamma\) are indicated by dashed lines.}\label{fig:hill}
\end{figure}

If a random variable has a power-law distribution, taking the logarithm of it transforms it into an exponentially distributed random variable, so another way of interpreting the Hill estimator is the following: it first transforms the power-law sample into an exponential one, and after that it estimates the parameter of this exponential by taking the sample average, using the fact that for \(X\sim\mathrm{EXP}(\lambda)\), \(\eval{X|X>l} = l + \lambda^{-1}\) for any \(l \in \mathbb R_+\), which is sometimes referred to as the ageless or memoryless property of the exponential distribution. A third interpretation was given in the context of generalized regression in \cite{Aban2004}.

The term Hill plot stands for plotting \(\hat \gamma^{\mathrm H}\) as a function of the tail length \(k\). If we know that the data is exactly power-law, it is of course best to set \(k = N - 1\), so the Hill plot does not make much sense. For real-life data, however, it is usually only the largest observations for which power-law behavior is a reasonable approximation. Consequently, one would normally look for a plateau in the Hill plot where the estimate is relatively stable, and choose the number of points to take into account based on that. Drees et al.\ \cite{Drees_HillPlot} have shown that it is best practice to plot the Hill estimator as a function of \(\ln k\) instead of \(k\), for which they coined the phrase altHill plot (figure \ref{fig:hill}). 

Taking a closer look, for a general regularly-varying distribution \eqref{eq:RV} the presence of the function \(L(x)\) complicates the situation in two ways:
\begin{itemize}
	\item It introduces a non-zero bias which becomes zero only asymptotically. How fast it becomes zero depends on the nature of corrections to the power-law behavior.
	\item The optimal tail length \(k_{\mathrm{opt}}\) which minimizes the mean squared error of the estimator becomes (often considerably) smaller than the sample size.   
\end{itemize}
See for example figure \ref{fig:hill}(c), showing the Hill plots for log-gamma distributed samples, which have a logarithmic correction to the power-law. Having information about \(L(x)\) is therefore very useful: one can modify the Hill estimator to reduce its bias, and also estimate the optimal tail length \(k_{\mathrm{opt}}\). The statistics literature contains several approaches to bias reduction as well as the choice of tail length (see section \ref{sec:tail_length} or \cite{Beirlant_StatisticsOfExtremes} and references therein), but in general, the more information one has about the higher-order behavior of the distribution the better, as estimating higher-order parameters is of course even more problematic than estimating the extreme value index \(\gamma\) itself. 

Another issue with the Hill estimator is that it definitely produces a positive number, regardless of the underlying distribution, see for example figure \ref{fig:hill}(b), where the true extreme value index \(\gamma\) is zero for the log-normal distribution. Generally, the closer the estimate \(\hat\gamma\) is to zero, the less confident one should be in the results, even if the histogram seems straight at the end. As a rule of thumb, if one has an estimate \(\hat\gamma < 0.3\), it is a good idea to consider alternative models for the data (unless the theoretical background is clear and indicates a power-law behavior without reason for doubt).

There are further, still basic tools, like the mean excess plot \cite{Coles2001, Ghosh2010}, which is quite popular in the actuarial field, however, but are not as well suited for estimating the value of the tail exponent.

\section{Parameter estimation for intensity measurements}\label{sec:intensity}

The previous section showed that in the ideal case, either approach yields results consistent with the true tail exponent. Actual experiments, however, are never that simple, the distortions and noise introduced by the apparatus require further considerations. There are two major effects that cannot be escaped: detector imperfections (limited linear response) and noise added by different experimental elements. We will use a very simplistic model of the experiments done in \cite{Manceau2019} in order to show how these affect estimation.

\subsection{Models used in simulations}\label{model}
Manceau \emph{et al.}\ \cite{Manceau2019} showed that for supercontinuum generation setups, the distribution of measured intensities has very heavy tails. Table \ref{tab:experiments} summarizes the types of experimental setups used by them from the mathematical point of view.
\begin{table}[htb]
\centering
\renewcommand{\arraystretch}{1.5}
\begin{tabular}{c|c|c|c|c}
Source & Experiment & Idealized distribution & Heavy-tailed? & \(\gamma\)
\\ \hline
Thermal & Optical harmonic gen. & \(\left[\mathrm{EXP}(\lambda)\right]^n\) & Y & 0 \\
Thermal & Supercontinuum gen. & \(\sinh^2\left[\mathrm{EXP}(\lambda)\right]\) & Y & \(2/\lambda\) \\
BSV & Optical harmonic gen. & \(\left[\mathrm{GAMMA}(1/2, \beta)\right]^n\) & Y & 0 \\
BSV & Supercontinuum gen. & \(\sinh^2\left[\mathrm{GAMMA}(1/2, \beta)\right]\) & Y & \(2/\beta\) \\
\end{tabular}
\caption{Experimental setups in \cite{Manceau2019}, BSV stands for bright squeezed vacuum. In effect we have different non-linear transformations of an exponential random variable for thermal pumping, and a gamma distributed variable for BSV pumping.}\label{tab:experiments}
\end{table} 

According to Equation \eqref{heavy_def}, each experimental setup produces heavy-tailed observables, even though the pumps are not heavy-tailed (the gamma distribution also decays asymptotically at an exponential rate). Note that for optical harmonic generation, even though the decay is slower than exponential, each moment exists, meaning that there is no theoretical issue with calculating second (or higher) order correlations like \(g^{(2)}\) \cite{Mandel1995}, so we are not going to further concern ourselves with that case. For the supercontinuum setup, however, the situation is quite different: depending on the value of \(\lambda\) (or \(\beta\)), which is inversely proportional to the pump's mean intensity, the highest existing moment can be arbitrarily small.

\begin{figure}[ptb]
\centering
\includegraphics[width=\textwidth]{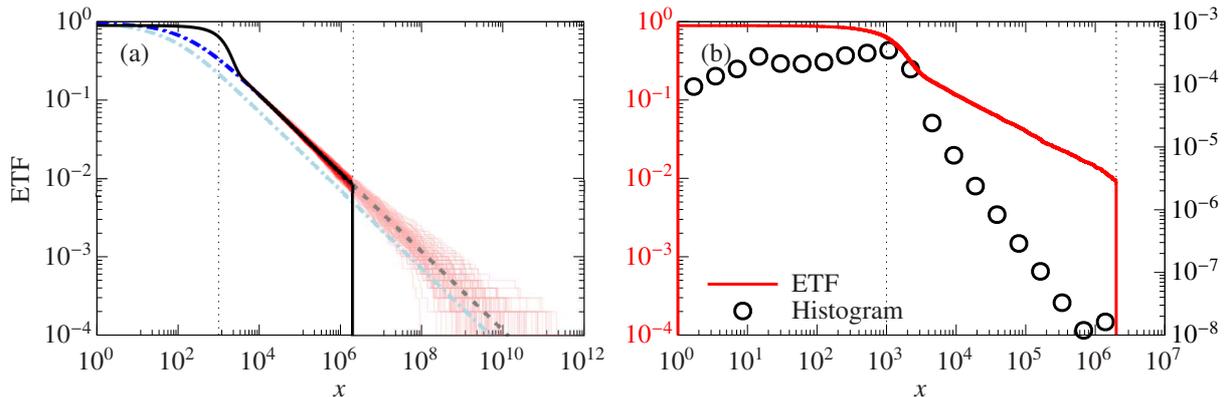}
\caption{a) Model behavior with thermal pumping. The light blue dash-dotted line shows the TF of \(X_0\), the dark blue dash-dotted line shows the TF of \(X_1\). The red/light red lines show 100 simulated ETF-s with lower cutoff and detector noise (\(X_2\)). The solid black line shows the TF of the model when an upper cutoff is also included. The lower and upper cutoffs are marked by black dotted lines. Parameters: \(I \sim \mathrm{EXP}(\lambda = 1)\), \(K = 200\), \(\omega_1\sim \mathcal N (0,  \sigma_1^2 = 1)\), \(l = 10^3\), \(\omega_2\sim \mathcal N (0, \sigma_2^2 = 10^6)\), \(L = 2\cdot 10^6\). (b) Histogram and ETF for a single sample of size $10^4$ with the given parameters.}\label{fig:modelbehavior}
\end{figure}

If the models in Table \ref{tab:experiments} were not only approximations of the reality, one would only need to take the inverse of the non-linear transformation involved in order to obtain a vanilla exponential or gamma sample and quite simply estimate the single free parameter of those distributions. Experimental imperfections do, however, complicate the situation. For illustration, let us consider the following simple model of the supercontinuum-generation process:
\begin{equation}
X = D\left(K\cdot \sinh^2(I + \omega_1)\right),
\end{equation}
with \(I\) denoting the pump's intensity (whose distribution is either \(\mathrm{EXP}(\lambda)\) or \(\mathrm{GAMMA}(1/2, b)\)) and \(X\) the detector's output. The detector is modeled through 
\begin{equation}
	D(x) = \left\{
	\begin{array}{ll}
		l + \omega_2 & \mathrm{if}\, x < l, \\
		x + \omega_2 & \mathrm{if}\, x \in [l, L], \\
		L + \omega_2 & \mathrm{if}\, x > L,
	\end{array}
	\right. 
\end{equation}
with \(\omega_1, \omega_2\) denoting independent Gaussian noises on the pump's and the detector's side, respectively (these can be aggregates of different types of noises); \(K\) is a constant factor. The value of \(l\) corresponds to the detector's noise floor, the value of \(L\) to its saturation limit. 

The effects of the different elements of the model are shown in figure \ref{fig:modelbehavior}(a). The dash-dotted light blue line shows the ideal case, i.e., the TF of \(X_0 \equiv K\cdot \sinh^2 I\) which is distorted in the following steps: Adding a Gaussian noise prior to the non-linear transformation (\(X_1 \equiv K\cdot \sinh^2 \left[I + \omega_1 \right]\)) corresponds asymptotically only to a multiplication by a constant factor \(\exp\left\{\lambda \sigma_1^2\right\}\) (see dark blue dash-dotted line). Introducing a lower cutoff \(l\) introduces a discontinuity in the TF, which is smeared by the second Gaussian noise \(\omega_2\) (\(X_2 \equiv \max\left\{K\cdot \sinh^2 \left[I + \omega_1 \right], l\right\} + \omega_2\)). The upper cutoff in \(L\) introduces a sharp fall to zero in \(L\), which could be smoothed by taking a more accurate model of the saturation curve. However, for the sake of simplicity, the final step is just \(X = \min\left\{\max\left\{K\cdot \sinh^2 \left[I + \omega_1 \right], l\right\}, L\right\} + \omega_2\). For the specific parameters used in figure \ref{fig:modelbehavior}, detector saturation essentially corresponds to discarding about the top 1\% of observations, shown in the figure in light red. Note that out of the parameters of the model, only the pump's mean intensity has an impact on the tail exponent (if there was a multiplicative distortion prior to the non-linear transformation, the situation would be different).  Figure \ref{fig:modelbehavior}(b) illustrates how much easier it is to notice if the saturation limit of the detector was reached during the experiment if one uses the ETF instead of the histogram to visualize the data: while there is a very visible cut-off at the saturation limit for the former, the latter shows only a slight increase in the rightmost bin. Note that in this model, there is a non-zero probability of negative observations, which is why the ETF is not go to one as \(x\) goes to zero.

In the following sections, we will examine how the different estimation tools perform for the different versions of our model.

\subsection{Numerical results for supercontinuum generation}\label{sec:SCG}
Using figure \ref{fig:modelbehavior}(a), we have gained an insight into what the individual parameters of our model do. Now we are looking at the performance of the three basic estimators discussed in section \ref{sec:tools}, and how they are affected by model parameters.

\begin{figure}[ptb]
\centering
\setlength{\unitlength}{0.5\textwidth}
\begin{picture}(2,0.7)
\put(0,0){\includegraphics[width=0.49\textwidth]{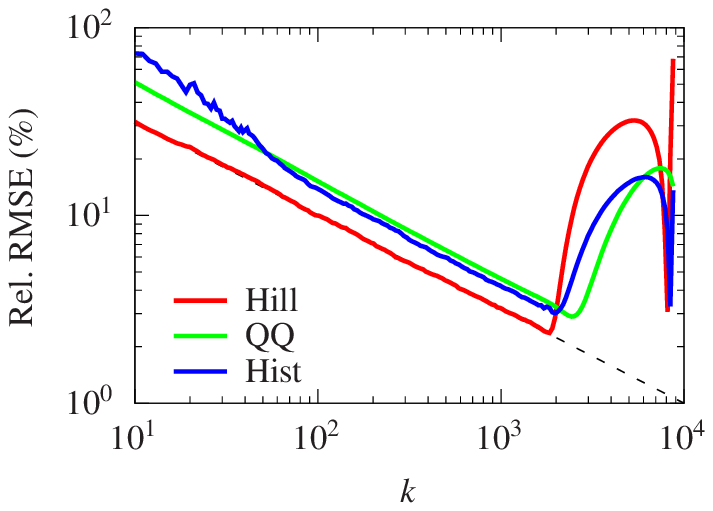}}
\put(0.85,0.6){(a)}
\put(1,0){\includegraphics[width=0.49\textwidth]{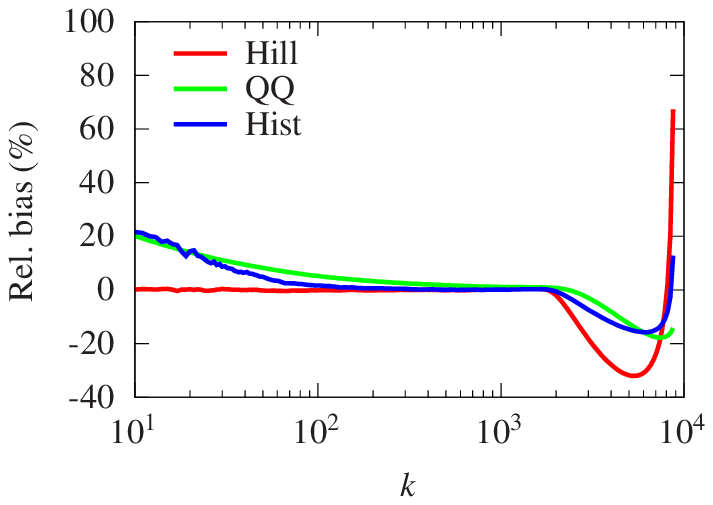}}
\put(1.85,0.6){(b)}
\end{picture}
\caption{Comparison of estimators, SCG setup, thermal source, without detector saturation. (a) Relative RMSE. The black dashed line shows the relative RMSE for the exact Pareto case, \(1/\sqrt{k}\). (b) Relative bias. (Parameters: \(\lambda = \sigma_1 = 1\), \(l = \sigma_2 = 10^3\), \(L = \infty\), \(K = 200\).)}\label{fig:cmp_exppump}
\end{figure}
Figure \ref{fig:cmp_exppump} shows the root mean square error of the different estimators for the thermal case, calculated using \(10^3\) simulated samples of size \(10^4\). Clearly, in this case the convergence to power-law is quick, since the ideal RMSE (black dashed line) is reached easily by the Hill estimator, with the other two estimators performing only a little worse (the minimum is about 4\% instead of 3\%). This is easily understood when examining the asymptotic form of the TF (which has a closed analytic form for \(\sigma_2 = 0\)),
\begin{equation}
\label{eq:thermal_asymptotics}
\overline F_{\mathrm{SCG, thermal}}(x) = \left[e^{-\lambda \sigma_1^2}\cdot  \frac{4x}{K}\right]^{-\frac \lambda 2}\cdot \left(1 - \frac {\lambda K}{4x} + \mathcal O(x^{-2})\right),
\end{equation}
which means that the convergence to the asymptotic behavior is also power-law. The divergence in the bias is caused by the fact that observations for this model can get arbitrarily close to zero (can even be negative), so the term \(-\ln x_{(k+1)}\) in \eqref{eq:hill} will dominate in the Hill estimator as \(k\) is increased. This divergence also introduces meaningless minima in the RMSE curve at \(k\approx 8 \cdot 10^3\), which is where the bias curves cross zero before diverging. 

\begin{figure}[t]
\centering
\setlength{\unitlength}{0.5\textwidth}
\begin{picture}(2,0.7)
\put(0,0){\includegraphics[width=0.49\textwidth]{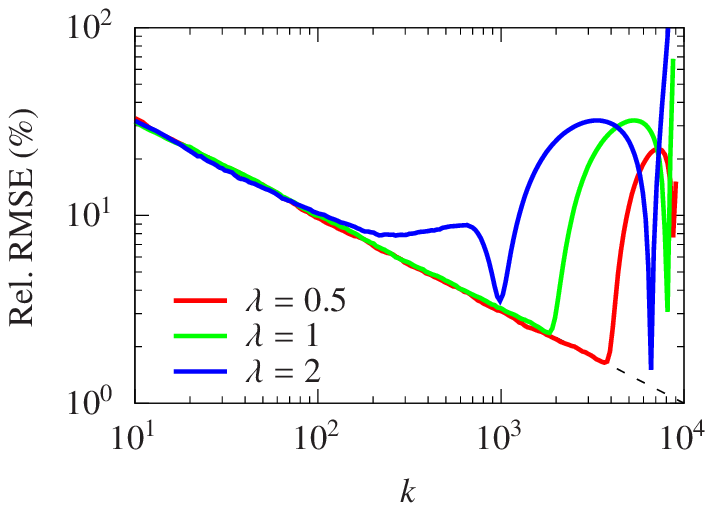}}
\put(0.85,0.6){(a)}
\put(1,0){\includegraphics[width=0.49\textwidth]{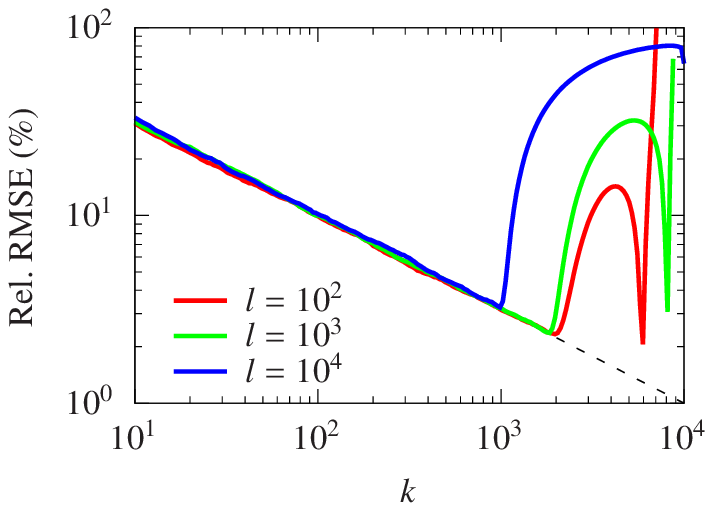}}
\put(1.85,0.6){(b)}
\end{picture}
\caption{Relative RMSE of the Hill estimator as a function of the number of points taken into account: (a) for different values of the parameter \(\lambda\) of the thermal pump;  (b) for different values of the noise floor parameter \(l\) of the detector.}\label{fig:exppump_var}
\end{figure}

Figure \ref{fig:exppump_var} shows that varying the mean intensity of the source (\(\lambda^{-1}\)) and the noise floor (\(l\)) changes the situation compared to figure \ref{fig:cmp_exppump}. The best achievable error is essentially determined by how many observations are unaffected by experimental imperfections. If the mean intensity is increased (figure \ref{fig:exppump_var}(a)), more observations are above the noise floor of the detector, which ultimately means that one has to discard fewer observations when estimating the tail index.  Note that the blue line in figure \ref{fig:exppump_var}(a) corresponds to considerable pre-transformation noise: the typical scale of the signal (\(\lambda^{-1} = 1/2\)) is exceeded by the scale of the noise (\(\sigma_1 = 1\)), resulting in oddly shaped profile. The same is true for the red line in figure \ref{fig:exppump_var}(b): even though the noise floor is low (\(l = 10^2\)), it is exceeded by the scale of the additive noise \(\omega_2\) (\(\sigma_2 = 10^3\)). 

\begin{figure}[t]
\centering
\setlength{\unitlength}{0.5\textwidth}
\begin{picture}(2,0.7)
\put(0,0){\includegraphics[width=0.49\textwidth]{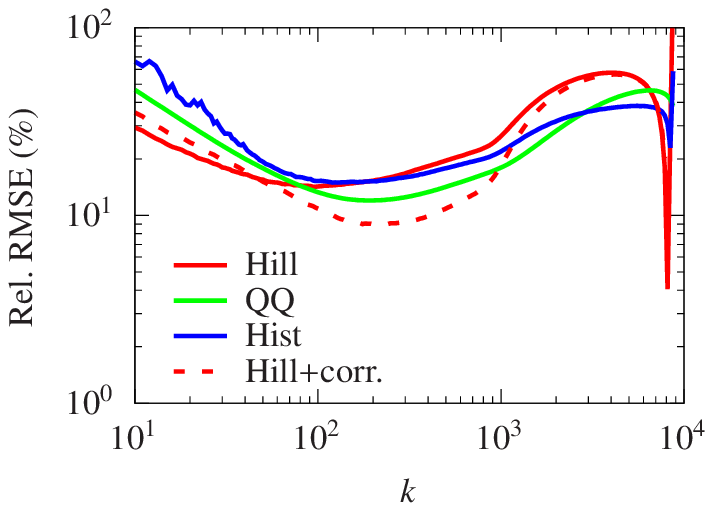}}
\put(0.85,0.6){(a)}
\put(1,0){\includegraphics[width=0.49\textwidth]{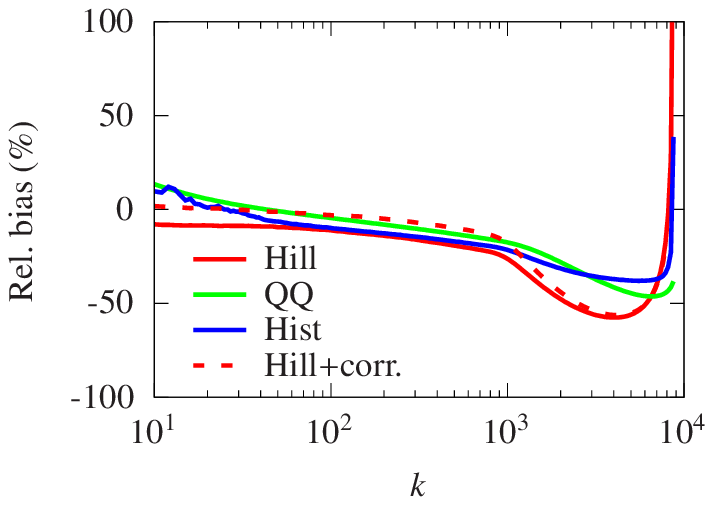}}
\put(1.85,0.6){(b)}
\end{picture}
\caption{Comparison of estimators, BSV source, without detector saturation. (a) Relative RMSE. (b) Relative bias. (Parameters: \(\beta = \sigma_1 = 1\), \(l = \sigma_2 = 10^3\), \(L = \infty\), \(K = 200\).)}\label{fig:cmp_gammapump}
\end{figure}

If the process is pumped by bright squeezed vacuum, the problem becomes more technically involved. As figure \ref{fig:cmp_gammapump} shows, the best achievable RMSE is significantly worse (10-20\% instead of 3-4\%), owing to the considerable bias of the estimators for this case. This is because the correction to power-law behavior decays as a power of \(\ln x\), and not \(x\). It is therefore helpful if, after taking the logarithm of the observations, one uses the conditional maximum likelihood estimator for \(\mathrm{GAMMA}(1/2, \beta)\) instead of the Hill estimator, which, as discussed before, is a conditional MLE for \(\mathrm{EXP}(\lambda)\). That is, one should maximize the function
\begin{eqnarray}
\nonumber \ln \mathcal L(\alpha, \beta) &=& \frac{\alpha - 1}{k}\sum_{i = 1}^k\ln\ln x_{(i)} - \frac \beta k \sum_{i = 1}^k \ln x_{(i)} \\
&&+ \alpha\ln \beta - \ln\Gamma(\alpha, \beta\ln x_{(k+1)}),
\label{eq:gamma_mle}
\end{eqnarray}
with \(\Gamma(s,z)\) denoting the upper incomplete gamma function. It is straightforward to show that asymptotically, the value of \(\beta^{-1}\) that maximizes \eqref{eq:gamma_mle} is equal to the Hill estimator, and that the corrections are \(\mathcal O\left(1/\ln x_{(k+1)}\right)\). The maximization can be done numerically using \(\alpha = 1/2 = \mathrm{fixed}\), with \(\beta_0 = \left[\hat \gamma^{\mathrm H}\right]^{-1}\) as the starting point. 
As the dashed lines in figure \ref{fig:cmp_gammapump} show, the estimator defined in \eqref{eq:gamma_mle} is indeed more efficient than the simple Hill estimator. The improvement is not as significant as one might hope for since the actual transformation applied to the input signal was \(K\cdot\sinh^2(\cdot)\) and not \(\exp(\cdot)\). Thus, in theory, to get better results, one should change \(\ln x_{(i)}\) to \(\mathop{\mathrm{asinh}}\left(\sqrt{x_{(i)}/K'}\right)\) in \eqref{eq:gamma_mle}, with \(K' = K\cdot \exp\left\{\sigma_1^2\beta\right\}\), however, this is problematic since \(K'\) depends on the unknown values of \(\beta\) (\(\equiv 2/\gamma\)), \(\sigma_1\), and \(K\).

\subsection{Detector saturation}\label{sec:saturation}
\begin{figure}
\centering
\setlength{\unitlength}{0.5\textwidth}
\begin{picture}(2,0.7)
\put(0,0){\includegraphics[width=0.49\textwidth]{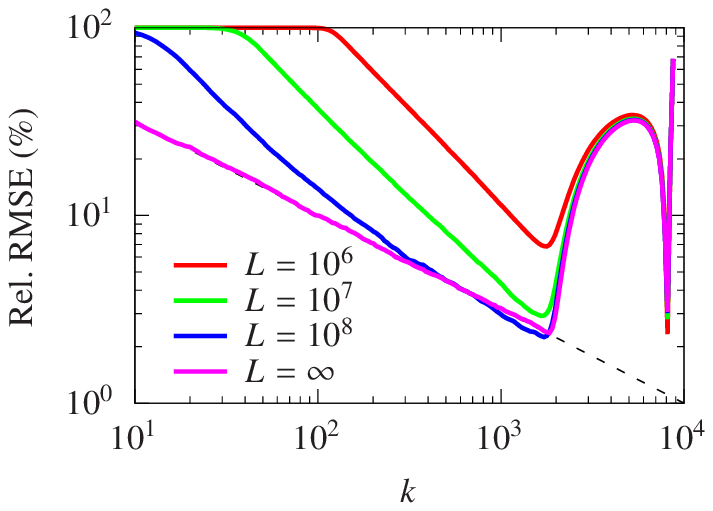}}
\put(0.85,0.6){(a)}
\put(1,0){\includegraphics[width=0.49\textwidth]{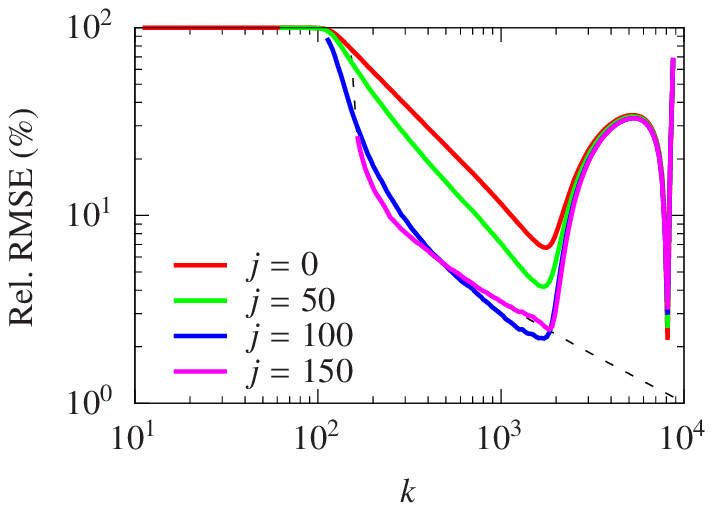}}
\put(1.85,0.6){(b)}
\end{picture}
\caption{(a) Hill estimator performance for different values of detector saturation. Dashed line: \(1/\sqrt k\). (b) Generalized version \eqref{eq:genhill} to compensate for the saturation \(L = 10^6\) and varying values of \(j\), the number of discarded observations. Dashed line: \(1/\sqrt{k - 150}\). }\label{fig:modifiedestimators}
\end{figure}

If the observations during an experiment are distorted by detector saturation, one can try and recalibrate the equipment so that all observations fall within the linear range of the detector. However, this is not necessarily trivial to do if the process is indeed heavy-tailed, but even if this is not a problem, it is still a good practice to try and evaluate the original data set instead of throwing it away. This, however, requires modifying the basic estimators introduced in section \ref{sec:tools}, which were constructed supposing that the largest observations are (close to being) Pareto distributed. Figure \ref{fig:modifiedestimators}(a) shows that, as expected, the basic approach fails if there is detector saturation.

The three basic techniques presented in section \ref{sec:tools} are easily modified to work with discarding the largest observations which are affected by detector saturation. For the least squares approaches on either the histogram or the QQ plot, one quite straightforwardly has to omit observations above a certain threshold, but otherwise the optimization is exactly the same as without the upper cutoff.

For generalizing the Hill estimator, one has to take advantage of the Rényi representation theorem according to which the scaled spacings of the order statistics of an exponential sample are themselves exponentially distributed: that is, if \(Z_i\), \(i = 1,\ldots, N\) is an i.i.d.\ \(\mathrm{EXP}(\lambda)\) sample, then \(S_{(i)} \equiv i\cdot(Z_{(i)} - Z_{(i+1)})\), \(i = 1,\ldots, N-1\) are also i.i.d.\ \(\mathrm{EXP}(\lambda)\), with \(Z_{(i)}\) denoting the \(i^{\mathrm{th}}\) largest observation in the sample. Knowing that if \(X\) is Pareto, then \(\ln X\) is exponential, discarding the \(j\) largest observations results in the following estimator:
\begin{eqnarray}
\nonumber \hat\gamma(k, j) &:=& \frac{1}{k - j} \sum_{i = j+1}^k i\left(\ln x_{(i)} - \ln x_{(i+1)}\right) \\
\label{eq:genhill}&=& \frac{j}{k-j}\ln \frac{x_{(j+1)}}{x_{(k+1)}} + \frac 1{k - j}\sum_{i = j+1}^k\ln \frac{x_{(i)}}{x_{(k+1)}}.
\end{eqnarray}
Note that setting \(j = 0\) does yield the standard Hill estimator, as expected. As figure \ref{fig:modifiedestimators}(b) shows, our suggested generalization indeed significantly improves the performance. When choosing the value of \(j\), one has to, of course, discard the values affected by detector saturation. However, discarding too many values increases the RMSE, which is in the best-case scenario proportional to \(1/\sqrt{k-j}\). If, in order to reduce the bias, one chooses to only keep the measurements within a shorter interval \(\left[x_{\mathrm{LO}}', x_{\mathrm{HI}}'\right]\) instead of \(\left[x_{\mathrm{LO}}, x_{\mathrm{HI}}\right]\) (\(x_{\mathrm{LO}} < x_{\mathrm{LO}}' < x_{\mathrm{HI}}' < x_{\mathrm{HI}}\)), the number of measurements has to be multiplied by a factor of  
\(
\left[{\overline F(x_{\mathrm{LO}}) - \overline F(x_{\mathrm{HI}})}\right] /
\left[{\overline F(x_{\mathrm{LO}}') - \overline F(x_{\mathrm{HI}}')}\right]
\).
This ensures that on average, the same number of observations will fall in \(\left[x_{\mathrm{LO}}', x_{\mathrm{HI}}'\right]\) during the second experiment as in \(\left[x_{\mathrm{LO}}, x_{\mathrm{HI}}\right]\) during the first one. The values of \(\overline F\left(\cdot\right)\) can be substituted by their empirical counterparts from the first experiment. This way one can check whether the new estimate \(\hat\gamma'\) gained from the second experiment is consistent with the old one. The figures for the QQ estimator are quite similar, whereas the histogram approach is not much improved by discarding the data affected by saturation.

\subsection{Choosing tail length}\label{sec:tail_length}
Figures \ref{fig:cmp_exppump} - \ref{fig:modifiedestimators} show the best attainable performance of the basic estimators. The problem in practice is how to decide how many points to take into account for calculating the estimators (\(k\)), especially if there is a large number of samples and evaluating each one is not feasible without automation. If the higher-order behavior is known, or can be estimated with reasonable accuracy, the optimal tail length (minimizing the RMSE) can be estimated as well. This is the approach followed by most of the literature, however, due to the difficulty of estimating higher-order behavior, the resulting algorithms can be quite involved (requiring the tuning of nuisance parameters) and often do not outperform simpler, heuristic approaches. 

\begin{table}[hptb]
\centering
\begin{footnotesize}
\renewcommand{\arraystretch}{1.1}
\begin{tabular}{|lccc|cc|cc|cc|}
\cline{5-10}
\multicolumn{4}{c|}{} & \multicolumn{2}{c|}{Pareto} & \multicolumn{2}{c|}{Log-gamma} & \multicolumn{2}{c|}{Model} \\
\hline
\multirow{2}{*}{Method} & \multirow{2}{*}{Based on} & \multirow{2}{*}{Tail choice} & \multirow{2}{*}{Cost} &
\multicolumn{2}{c|}{\(\gamma\)} & \multicolumn{2}{c|}{\(\gamma\)} & \multicolumn{2}{c|}{\(\gamma\)} \\ \cline{5-10}
& & & & 2 & 1 & 2 & 1 & 2 & 1 \\
\hline\hline
1. & Hill & path stability & mid &
\cellcolor{green!25}0.04 & \cellcolor{green!50}0.02 &
 0.46 & 0.22 &
 \cellcolor{green!75}0.08 & \cellcolor{green!75}0.15 \\

2. &  hist. & path stability & mid &
0.16 & 0.09 &
0.45 & \cellcolor{green!50}0.20 & 
0.27 & \cellcolor{green!25}0.22 \\

3. &  QQ & path stability & mid &
0.30 & 0.16 &
0.52 & 0.23 & 
\cellcolor{green!25}0.24 & \cellcolor{green!25}0.22 \\

4. \cite{Clauset2009} &  Hill & KS distance & high &
\cellcolor{green!50}0.03 & \cellcolor{green!75}0.01 & 
0.46 & 0.23 & 
\cellcolor{green!75}0.08 & 0.44 \\

5. \cite{Guillou2001} &  Hill & aux.\ statistic & low &
\cellcolor{green!50}0.03 & \cellcolor{green!50}0.02 &
\cellcolor{green!75}0.39 & \cellcolor{green!75}0.18 & 
\cellcolor{green!50}0.13 & \cellcolor{green!50}0.17 \\

6. \cite{Danielsson2001} &  Hill & higher-order p. & high &
\cellcolor{green!75} 0.02 & \cellcolor{green!75}0.01 &
\cellcolor{green!50} 0.40 & \cellcolor{green!50}0.20 &
 0.64 & 0.56\\

7. \cite{Drees1998} & Hill & higher-order p. & mid &
0.15 & 0.09 &
0.46 & 0.23 &
0.25 & 0.61 \\

8. \cite{Caeiro2015}/1 & Hill & higher-order p. & mid &
0.05 & \cellcolor{green!25}0.03 &
0.43 & \cellcolor{green!25}0.21 &
0.81 & 0.37\\

9. \cite{Caeiro2015}/3 &  Hill & higher-order p. & high &
0.07 & 0.04 &
0.46 & 0.23 & 
0.89 & 0.63 \\

10. \cite{Caeiro2015}/3 & reduced-bias Hill & higher-order p. & high &
0.21 & 0.11 & 
\cellcolor{green!25}0.41 & \cellcolor{green!50}0.20 &
 0.90 & 0.72 \\
 \hline
\end{tabular}
\end{footnotesize}
\caption{Empirical RMSE of different estimators applied to 100 samples of size $10^4$, for different types of distributions, all of which had a positive finite extreme value index ($\gamma \equiv \alpha^{-1} \in \{2, 1\}$). The ``Cost'' column refers to CPU time and complexity. The column ``Model'' refers to the toy model with a thermal pump with (\(\lambda=\gamma/2\), \(\sigma_1 = 0\), \(l = \sigma_2 = 10^3\), \(L = \infty\)). The best three values were indicated in different shades of green for each distribution. \mytodo{Make more compact}}\label{tab:rmse}
\end{table}

We have implemented the estimation procedures shown in Table \ref{tab:rmse}. We proposed procedures 1-3 in which a heuristic path stability approach was used to choose the tail length, mainly based on \cite{Neves2015}, and tweaked using \cite{Drees_HillPlot} and \cite{Clauset2009} (for details, see \ref{sec:PS}). This approach was included because it essentially emulates how one would choose a tail length from the Hill plot, and it also works with an arbitrary estimator. Table \ref{tab:rmse} of course, is not an exhaustive study of these procedures, especially since procedures 6-10 involve more than one nuisance parameter, which could have been used to fine-tune them. We did not do that, we used the default parameter values suggested in the sources to see how they perform ``out of the box''. Based on our tests, the procedure introduced by Guillou et al.\ \cite{Guillou2001} provided the most reliable results and had the further advantage of simplicity and speed. Our suggested path stability approach, which mimics visual evaluation, also works reasonably well combined with the Hill estimator.

\section{Conclusion and Discussion}
In this work, we first gave a brief overview of a basic toolkit that may be used to estimate the tail exponent of associated with a heavy-tailed sample. We devoted more attention to histograms than it is justified from the mathematical point of view as they are the default tools for many physicists, and discussed how to use them more efficiently. Subsequently, we discussed two simple alternatives: the QQ estimator and the Hill estimator through examples. 

We addressed the challenges specific to intensity measurements in supercontinuum generation experiments. In order to do that, we introduced a model for the observable intensity distribution. Firstly, if the source in the experiment is bright squeezed vacuum, the estimation becomes considerably more inefficient than in the thermal case. We suggest using a modified version of the Hill estimator, defined in \eqref{eq:gamma_mle} for BSV. If the pump is thermal the plain Hill estimator \eqref{eq:hill} is sufficient. Next, one should check whether the observations were affected by detector saturation. This is easily done by preparing the empirical tail function of the sample. 
If the answer is yes, one has to discard the affected observations and apply the estimator introduced in \eqref{eq:genhill}. 
Finally, we also included comparison of procedures which can choose how many observations to take into account in an automated fashion. 

These results can be directly extended to investigate heavy tail distribution caused by three-wave and four-wave mixing in non-linear optics \cite{Manceau2019}, atomic physics \cite{Boyer2008}, superconducting circuits \cite{Sivak2019}, and non-linear optomechanics \cite{Brawley2016, Siler2018}, and electromechanics \cite{Seitner2017}. 
\section*{Acknowledgments} 
{\'E}.R. and L.R. acknowledge the support of project 19-22950Y of the Czech Science Foundation.
R.F. acknowledges the projects {LTAUSA19099} and {CZ.02.1.01/0.0/0.0/16{\_}026/0008460} of MEYS CR.

\bibliography{heavy_tailed.bib}{}
\bibliographystyle{ieeetr}

\appendix
\vspace{0.5cm}
\begin{flushleft}
{\bfseries \Large Appendix}
\end{flushleft}

\section{Path stability approach to choosing tail length}\label{sec:PS}
\begin{figure}[ptb]
\centering
\includegraphics[width=0.7\columnwidth]{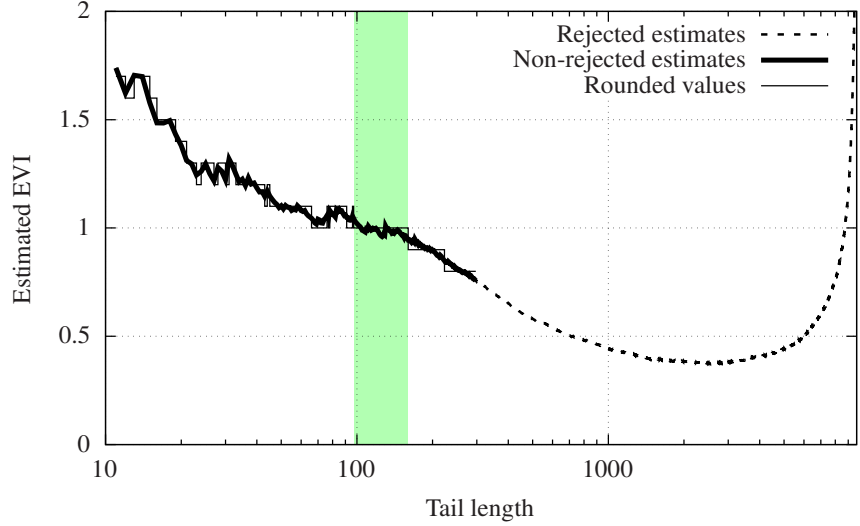}
\caption{Heuristic tail length selection algorithm. The thick black line corresponds to \(\mathcal K\), the green interval to \(\mathcal I\). The sample was created by taking applying \(\sinh ^2(\cdot)\) to an exponential sample and adding a Gaussian noise. The true value of \(\gamma\) for this sample is one. }\label{fig:alg}
\end{figure}

Procedures 1-3 in Table \ref{tab:rmse} use the following algorithm, which was adapted from \cite{Neves2015}, to select the tail length: 
\begin{enumerate}
\item Calculate the value of the estimator for all applicable values of the tail length: \(\hat \gamma(k)\) for \(k=k_{\min},\ldots,k_{\max}\).
\item For all \(k_{\min} \leq k \leq k_{\max}\), perform a statistical test to see whether a power-law with the estimated parameters can be rejected or not. Let \(\mathcal K\) denote the uninterrupted interval of values of \(k\) which does not include any rejected tail lengths and is to the leftmost. See the dashed portion in Figure \ref{fig:alg} showing the rejected values. This addition was inspired by \cite{Clauset2009}, although the KS statistic itself is used in a much different manner.
\item For \(\forall k\in \mathcal K\), round the estimated value \(\hat{\gamma}(k)\) to \(n\) decimal places. Choosing \(n = 1\) is mostly appropriate, the important thing is that the rounded values should not be constant within \(\mathcal K\). 
\item Identify the longest interval \(\mathcal I\), within which the rounded estimates are constant. The original method suggests defining length as the number of elements (\(k_2 - k_1 + 1\)), we suggest per Drees et al.\ \cite{Drees_HillPlot} defining length on the log scale (\(\log (k_2/k_1)\)). 
\item Within \(\mathcal I\), calculate the estimates rounded to \(n+2\) digits, and identify the mode of these values, denoted by \(m\).
\item The estimate of the tail is then gained by choosing the shortest tail \(\hat k\) within \(\mathcal I\) for which \(\hat\gamma(\hat k)\) rounded to \(n+2\) decimal places is equal to \(m\). The final estimate of \(\gamma\) is then \(\hat\gamma (\hat k)\).

\end{enumerate}
%

%
%
%

\end{document}